\documentclass[10pt,conference]{IEEEtran}
\IEEEoverridecommandlockouts
\usepackage{cite}
\usepackage{amsmath,amssymb,amsfonts}
\usepackage{algorithmic}
\usepackage{graphicx}
\usepackage{textcomp}
\usepackage{tabularx}
\usepackage{colortbl}
\usepackage{balance}
\usepackage{longtable}
\usepackage{caption}
\usepackage{subcaption}
\usepackage[nolist]{acronym}
\usepackage[draft]{hyperref}
\usepackage{xcolor}
\usepackage{todonotes}
\def\BibTeX{{\rm B\kern-.05em{\sc i\kern-.025em b}\kern-.08em
    T\kern-.1667em\lower.7ex\hbox{E}\kern-.125emX}}

\def\IEEEauthorrefmark#1{\raisebox{0pt}[0pt][0pt]{\textsuperscript{\footnotesize\ensuremath{\ifcase#1\or *\or \dagger\or \ddagger\or%
				\bullet\or \circ\or \cdot\or \times\or \checkmark\or **\or \dagger\dagger%
				\or \ddagger\ddagger \else\textsuperscript{\expandafter\romannumeral#1}\fi}}}}

\author{
	\IEEEauthorblockN{Eric Lanfer\IEEEauthorrefmark{1}, Thomas Hänel\IEEEauthorrefmark{1}, Roland van Rijswijk-Deij\IEEEauthorrefmark{5}, Nils Aschenbruck\IEEEauthorrefmark{1}}
	\vspace*{.18cm}
	\IEEEauthorblockA{
		\begin{tabular}{c c}
			{\IEEEauthorrefmark{1}Osnabrück University}&{\IEEEauthorrefmark{5}University of Twente} \\
			 {Institute of Computer Science}&{Design and Analysis of Communication Systems Group}\\
			{Osnabrück, Germany} &{Enschede, The Netherlands}   \\
			\small\texttt{\{lanfer, haenel, aschenbruck\}@uos.de} & \texttt{r.m.vanrijswijk@utwente.nl}
		\end{tabular}
}}

\title{Improving Proximity Classification for Contact Tracing using a Multi-channel Approach}
\begin{document}

\maketitle

\begin{abstract}
Due to the COVID 19 pandemic, smartphone-based proximity tracing systems became of utmost interest. Many of these systems use \acf{BLE} signals to estimate the distance between two persons. The quality of this method depends on many factors and, therefore, does not always deliver accurate results. In this paper, we present a multi-channel approach to improve proximity classification, and a novel, publicly available data set that contains matched IEEE~802.11 (2.4~GHz and 5~GHz) and \ac{BLE} signal strength data, measured in four different environments. We have developed and evaluated a combined classification model based on \ac{BLE} and IEEE~802.11 signals. Our approach significantly improves the distance classification and consequently also the contact tracing accuracy. We are able to achieve good results with our approach in everyday public transport scenarios. However, in our implementation based on IEEE~802.11 probe requests, we also encountered privacy problems and limitations due to the consistency and interval at which such probes are sent. We discuss these limitations and sketch how our approach could be improved to make it suitable for real-world deployment.
\end{abstract}

\section{Introduction}
	The scale and impact of the COVID-19 pandemic spurred governments and society to look for easy ways to automatically track contacts between individuals in order to trace chains of infection. Several countries and technology companies came up with software-based solutions implemented on smartphones. One such approach, which is deployed by many countries, is \ac{GAEN} \cite{Exposure:online}. This method is based on logging potential contacts using \acf{BLE} advertisements. This approach to contact tracing is used by many countries and implemented as mobile applications by the national and federal health agencies. Well-known examples are Germany, Ireland, Italy, the Netherlands, South Africa, and several states in the US.
	
	The risk of exposure is assessed using a distance approximation based on signal strength information (RSSI and TX Power). Such distance approximations tend to be noisy and error-prone \cite{8444603,MADOERY2021101474}. Studies, e.g., by Leith and Farrell \cite{leith2020measurement} or the calibration trial for the Singaporean contact-tracing app \cite{opentrac83:online} show how challenging this approach can be. There are many factors that can affect the signal strength of a \ac{BLE} signal, for example antenna design, Bluetooth stack implementation, or sending power. Equally, the quality of classifications can be influenced by the environment through common factors that always exist when it comes to radio wave propagation: attenuation, shadowing, reflection, scattering, diffraction, and refraction.
	Due to these limitations, almost no infectious contacts were detected in \cite{leith2020measurement}. Even if some were detected, the results showed high false-positive rates.

	To deliver on the promise of automated contact tracing, it is important that the accuracy of distance classification is improved. We argue that the combination of IEEE~802.11 broadcast messages (2.4~GHz and 5~GHz) with \ac{BLE} advertisements will significantly improve the classification of the proximity between two devices. Our core contributions of this paper are as follows:	
	\begin{itemize}
		\item We introduce a multi-channel proximity classification technique for contact tracing applications based on a combination of \ac{BLE} with IEEE 802.11 probe requests in the 2.4 and 5 GHz bands.
		\item We evaluate our approach in four different scenarios (outdoor, office, and two public transport settings on a bus and a train) and make our measurement results available as open access data.
		\item We show that our approach significantly outperforms \ac{GAEN} in terms of accuracy of distance classification.
	\end{itemize}
	Moreover, we evaluate several machine learning models for the use case of proximity classification. Beside this use case, our approach could also be used in other fields where accurate proximity classification is a major issue, such as localization, indoor navigation, or beaconing products. Finally, we provide suggestions on how our approach can be incorporated into existing contact tracing solutions to reduce the number of false-positives and false-negatives. 

	The remainder of this paper is structured as follows. Section~\ref{section:background} provides background information on wireless proximity classification and current approaches to contact tracing. Next, Section~\ref{section:relatedwork} discusses related work, followed by a detailed discussion of our approach in Section~\ref{section:approach}. In Section~\ref{section:dataset}, we provide information on the data set we recorded and used for evaluation. We continue in Section~\ref{section:evaluation} by presenting the evaluation of our approach, and end with conclusions and future work in Section~\ref{section:conclusion}.
	
\section{Background}
\label{section:background}

\subsection{Bluetooth Low Energy Advertisements}
\acf{BLE} was introduced in Bluetooth specification 4.0. It includes advertisements, which are broadcast messages that are delivered to all clients in signal range. It was originally developed
for Internet of Things (IoT) use cases, for example to deliver sensor data energy
efficient between several sensor nodes, or to implement beacons for localization purposes. Being available in almost every smartphone, these advertisements were then quickly utilized in contact tracing solutions. Most of the contact tracing frameworks send out messages with randomized advertising addresses (\emph{AdvA}) \cite{CoreSpec16:online}. 
The advertisements are only sent on 3~of the 40~\ac{BLE} channels, the three ADV\_CHs are channel 37 (2402 MHz), channel 38 (2426 MHz), and channel 39 (2480 MHz). These three frequencies were selected in an attempt to minimize interference with IEEE 802.11 b/g/n channels 1, 6, and 11. Even though the ADV\_CHs have been chosen to have low interference rates, studies show that channel 37 and 38 are still significantly affected by noise and interference from IEEE 802.11 \cite{nikoukar2018empirical}. This causes variation in signal strength, which is later used to determine the proximity between two devices.

\subsection{\acf{GAEN}}
	The \acf{GAEN} framework is used by at least 15 EU member states and several other countries and regions around the world to implement their proximity tracing solution \cite{martin2020demystifying}. The \ac{GAEN} concept derives from the \ac{DP-3T} proposal, which was presented after a wide discussion on privacy in contact tracing frameworks \cite{troncoso2020decentralized}. A \ac{BLE} advertisement sent out by \ac{GAEN} carries its information in the payload field. A 16 byte \ac{RPI} is used to warn a user, followed by 4 bytes of associated encrypted metadata. The metadata contains information of the framework version and the transmission power (TX\_power) level. The \ac{RPI}s are used to detect potential contacts. When a potential match is found, the metadata gets decrypted and a risk assessment will be done by estimating the distance to the infectious individual and the duration of a possible exposure \cite{Exposure17:online}. To estimate the distance between the two individuals, this formula is used: 	
	\begin{equation}
		\text{Attenuation}=\text{TX\_pwr}-(\text{RSSI\_measured}+\text{RSSI\_corr})
	\end{equation}
	
	In this formula $\text{TX\_power}$ is obtained from the metadata, $\text{RSSI\_measured}$ from the local log, and $\text{RSSI\_correction}$ from a vendor-supplied list of correction values for various smartphone models. Since version $1.5$ of the API and on smartphones with Android~9 or higher, for each scan window the minimum and average attenuation values are available. Due to the fact that the three ADV\_CHs may have different noise levels, that result in various \ac{RSSI}, it is recommended to use the minimum attenuation for further risk assessment, as this reflects the best signal transmission situation over the three channels.

	The next step depends on the implementation of the actual proximity tracing app, as different thresholds for calculations are used. In essence, the framework calculates a weighted risk score using these thresholds. It first classifies the attenuation as \emph{very close}, for distances less than 1.5m, with an attenuation value less than $d_{vc}$~dB, \emph{close}, for distances less than 3m, with an attenuation value less than $d_c$~dB, or \emph{safe} distances for all higher values. These classified attenuation durations can also be obtained by querying the API. As an example, the thresholds in the German Corona Warn App (CWA) are $d_{vc}=55$~dB for very close and $d_c=63$~dB for close, with any values higher than that classified as safe. 
	
	Next, the time is added up for the two close cases using the following formula:
	\begin{equation}
		ES=B1+0.5B2
	\end{equation}	
	In this formula, $B1$ is the time of exposure in the \emph{very close} category, so less than 1.5m and $B2$ the time of exposure at \emph{close} distance, so less than 3m. If the resulting value is at least $15$, the user will be warned by the application and informed about further steps to take \textit{(e.g., getting a test etc.)} \cite{cwadocum66:online}.  The attenuation thresholds vary, depending on the country-specific implementation by the health authorities. 

	\subsection{IEEE 802.11 probe requests}
	IEEE~802.11 probe requests are messages that are sent out via broadcast, in order to detect which IEEE~802.11 networks are around. Depending on the device, the messages are sent out with a randomized MAC address via 2.4~GHz and/or 5~GHz periodically, when the devices are not connected to a network \cite{802.11std}. Many smartphones send these messages when they have \emph{screen-on-time}, e.g., when a user is interacting with the phone, in order to connect to a network to save on cellular data. By entering a network interface into monitor mode, it is also possible to capture these packets on conventional client hardware. When the packets are captured, the \ac{RSSI} can be captured and information on which specific frequency a packet was sent is also available \cite{freudiger2015talkative}. 
	
	Compared to \ac{BLE} advertisements, where it is unclear on the receiving side which ADV\_CH was used, this information being available for IEEE 802.11 broadcasts is a significant benefit, since more specific path-loss-models can be computed if the actual frequency is known~\cite{gentner2020identifying}. 
\subsection{Signal propagation}
	\label{section:signalprop}
	The transmission of radio signals is affected by several factors, that make it hard to model the propagation of a radio signal. These factors have strong influence on the power with which a signal is received. The energy lost on the transmission path is called \emph{path loss}. The most general issue of path loss is attenuation. When a signal is sent through a medium, it loses power proportional to the distance that the signal needs to travel through a medium. Signal propagation, especially indoors, is mainly affected by four effects \cite{bensky2019short,blaunstein2014radio}: Reflection, diffraction, scattering and multi-path propagation.
	
\begin{figure}[tb]
		\centering
		\includegraphics[width=0.83\linewidth]{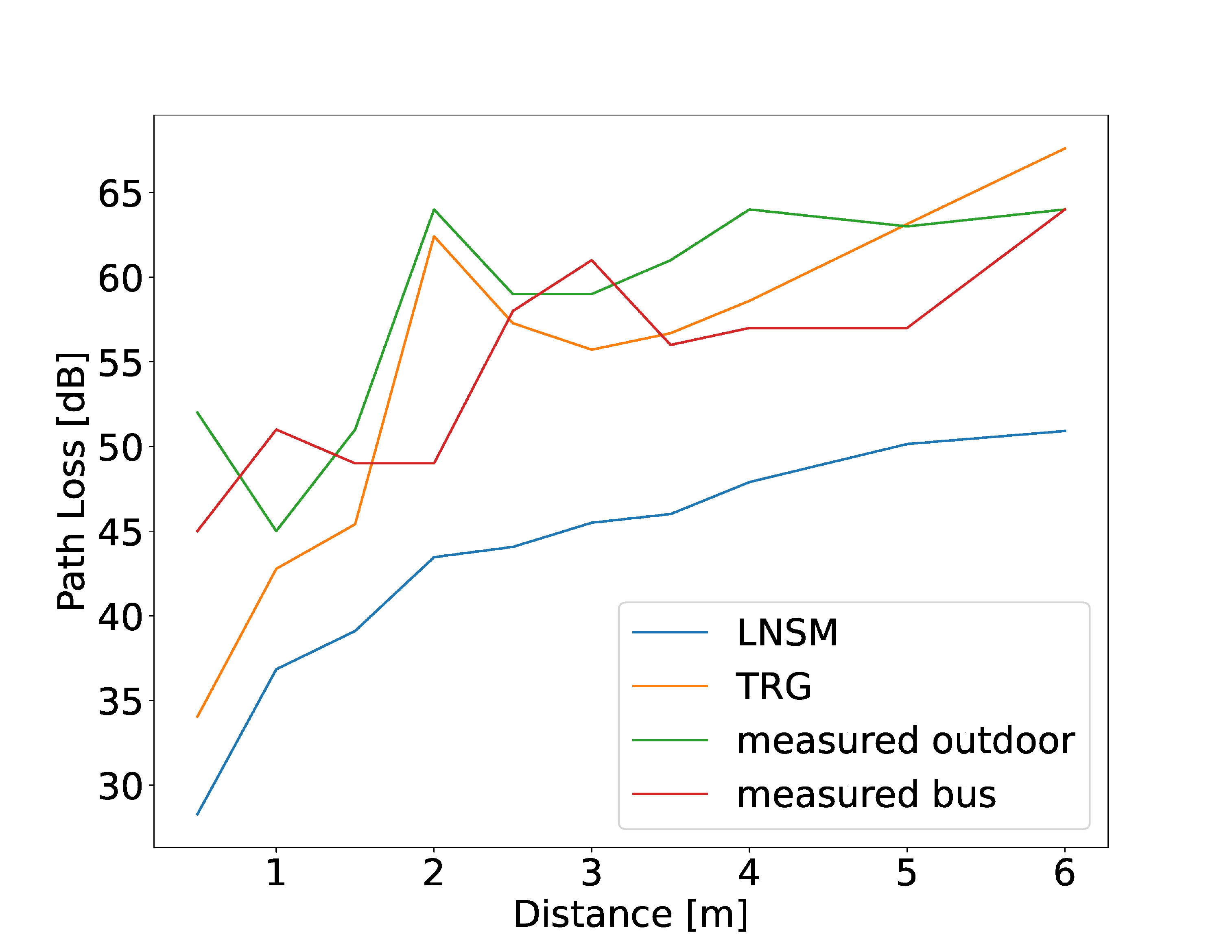}		
		\caption{Modelled vs. measured attenuation in indoor and outdoor scenarios using BLE channel 39 (2480 MHz). Modelled with log-normal-shadowing-model (LNSM) and two-ray-ground model (TRG)}
		\label{signal_modelling_ble}
	\end{figure}

	These effects make it hard to model signal strength according to the distance and to derive a distance from a given signal strength. Models such as the two-ray-ground model take factors like the ground reflection into account. For varying situations and environments, with unknown height levels and orientations of receiver and transmitter, as well as low information grades as we have for BLE advertisements (no frequency information), it is almost impossible to obtain an accurate model for estimating a distance. To illustrate this issue, we combined measured data in an indoor and an outdoor environment in a plot, together with the log-normal-shadowing-model and the two-ray-ground model, in Figure~\ref{signal_modelling_ble}. 

\section{Related work}
\label{section:relatedwork}
\subsection{Evaluation studies on proximity tracing}
One of the first studies, after GAEN was announced, was carried out by Leith and Farell \cite{farrell_bus,leith2020measurement}. They deployed smartphones in a tram and in a bus, placed the devices at various distances and collected GAEN advertisements and let them run through the risk assessment process. The results were computed for the German, Swiss, and Italian attenuation thresholds. Almost no alerts were triggered using these thresholds, even though the devices were at a critical distance. In later work \cite{leith2020coronavirus}, they assessed the correlation of RSSI and distance with a comparison of various environments. The authors pointed out that device orientation, human bodies in the \ac{LOS}, or the position of the device, e.g., whether it is in a pocket, play an important role and have crucial influence on the RSSI. In outdoor scenarios, a correlation can be seen, but in indoor scenarios too many factors come in, which make it very challenging to derive a correlation.

Zhao et al.~\cite{zhao2020accuracy} evaluated a set of proximity tracing applications, surveyed their broadcasting behaviour, the underlying calculations for proximity estimation, and tuning possibilities to the \ac{RSSI} measurements. They showed that there are different internal factors in software and in hardware that can affect the RSSI. For example, chipset and antenna of the smartphone, or the operating system running on a smartphone that is adjusting the levels of transmission power. Moreover, they briefly mentioned issues of signal propagation like obstacles and interference of other signals.

Since device-to-device proximity tracing was not deeply researched before the pandemic, there is little prior art in this space, especially when it comes to combining multiple channels. In the field of indoor navigation and positioning, however, many studies show how challenging it is to perform accurate proximity classification, especially in indoor environments. Even with complex preprocessing pipelines and sophisticated approaches for the removal of propagation effects, these approaches achieved results that still had an error of at least $1m$ \cite{vcabarkapa2015comparative,chen10.1145/3143361.3143385}. Approaches that showed better results were only fitted to a single environment, e.g., an office \cite{7042942,Palaghias7248384}. Such approaches are not applicable for contact tracing purposes, as this requires the approaches to work in a variety of environments in real-life scenarios and environments.

A number of studies evaluated the use of IEEE 802.11 signals in a device-to-device setup, these studies showed that IEEE 802.11 2.4~GHz signals can achieve more accurate results than \ac{BLE} signals \cite{7759023,6197489}. Additionally, a study by Ter\'{a}n et al.~\cite{7759023} evaluated the combination of \ac{BLE} and IEEE~802.11 2.4~GHz signals in a kNN classifier. They were able to obtain an accuracy of $0.75$ in a resolution of $1m$, which was a better result than using a single signal type in isolation. 

Regarding the specifics in signal propagation of the three \ac{BLE} ADV\_CHs, Nikoukar et al.~\cite{nikoukar2018empirical} collected data in various indoor and outdoor scenarios. In line with the other results, a trend of RSS against distance can be obtained from outdoor scenarios, but it is quite hard for indoor applications. They explained these with reflections, multi-path effects, interference by IEEE~802.11 networks inside, and antenna anisotropy. Moreover, they state that among these three channels, channel 39 is facing less interference and the measured values have less variance due to smaller IEEE~802.11 interference.

\subsection{Approaches improving proximity tracing}
Various studies try to improve \ac{BLE}-only proximity classification approaches. The NIST issued a contest for AI researchers to develop approaches to improve the accuracies of currently used proximity tracing products. The winners of this contest stated that it is very challenging to derive a distance from the \ac{RSSI} of a \ac{BLE} signal. Therefore, they utilized other sensors from a smartphone, such as \ac{IMU} and magnetron, in order to detect the environment and the orientation of a device~\cite{he20202}. The proposed neural networks have a high computational complexity and have to be trained for each device type. Therefore, the approach is not directly applicable for a wide deployment.	

Besides the NIST challenge, Clark et al. \cite{CLARK2021100168} suggested the usage of a network localization algorithm. This approach showed good results, but it requires a central instance that collects all RSSI values received from all devices to compute the algorithm, which will result in a privacy issue. 

Gentner et al.~\cite{gentner2020identifying} try to determine on which of the three channels the packet was received. The proposed approach is timing-based and runs on the application level in Android. Their results showed an accuracy of about $100\%$ in determining the correct \ac{BLE} channel. This information makes it possible to use more specific models, but still, this information does not help to model propagation issues.
	
\section{Approach}
\label{section:approach}
The main question that we want to answer in this paper is: \emph{Is it possible to improve the accuracy of proximity classification?} To answer this question, we hypothesize that improvement is possible by using a multi-channel approach, combining IEEE~802.11 and \ac{BLE} signals. Our assumption for this hypothesis is, that IEEE~802.11 signals have different signal propagation characteristics. Even for both signal types on 2.4~GHz, the IEEE~802.11 signals on 2.4~GHz are sent with higher transmission power and using different modulation methods. Moreover, these signals, especially on 5~GHz band, tend to suffer less from interference by noise. Having multiple \ac{RSSI} values for more than one signal type, and having exact frequency information from the IEEE 802.11 signals results in an information gain that should enable a classifier to make more accurate distance classifications between two smartphones. 

In order to prove that our hypothesis is correct, we present a novel data set of \ac{BLE} and IEEE~802.11 signals captured in various settings, at a variety of distances, and with three different devices in Section~\ref{section:dataset}. In our data set we trigger regular sending of IEEE~802.11 probe requests. This technique is already available and current proximity tracing deployments could be extended, in a way that also IEEE~802.11 probe requests are sent and captured for proximity classification. We use this data set to train various machine learning models and evaluate these models with typical machine learning metrics. The models predict, whether a signal was sent in \emph{very close} ($d < 1.5$~m), \emph{close} ($d \leq 3$~m) or \emph{safe} ($d > 3$~m) distance.

At first, we evaluate each signal type separately. Our goal is to determine the differences and the contributions of a single signal type to a distance classification. Especially, the \ac{BLE} threshold based approach is assessed here to provide an empirical evaluation of the classification quality. Due to the fact, that we use machine learning approaches for the other signal types, we also do train two models on \ac{BLE} signals only, to have an adequate comparison. Otherwise, advantages or disadvantages may have a bias due to the different classification approach.

Secondly, we create models that take all three signal types into account. We follow two different approaches:
\begin{enumerate}
	\item A combination of specialized classifiers that utilize a single signal type to predict the proximity. Thus, three models (\ac{BLE}, 802.11 2.4~GHz band, and 5~GHz band), get their specific input and predict the distance class. Then, all three results are combined to an overall result.
	\item One general classifier to build a model that utilizes all features (RSSIs of all three signals and frequencies for IEEE 802.11) directly as an input to perform a classification.
\end{enumerate} 


Overall, we test 13 models with different features (RSSIs and frequencies) for our approach. All 13 models are listed in Table~\ref{tbl:models}. Besides the threshold based \ac{BLE} classification, we also evaluate Decision Tree and Random Forest classifiers, implemented using \emph{sckit-learn}\footnote{\url{https://scikit-learn.org/}}, for certain signal types. We decided to use only Decision Tree-based classifiers and Random Forest, as a more sophisticated combination of trees in a first step. Our goal is to test if there are advantages in general of our multi-channel approach, due to the high explainability of the results. Moreover, it is possible to derive thresholds for the new signal types, from  the highest nodes of a Decision Tree. Having this, it would be possible to follow the same approach as GAEN uses for BLE also for the IEEE 802.11 features. The parametrization of the models was done using grid search. 

\begin{table}[t]
	\centering
	\caption{Models used for evaluation, with single signal type models 1-7, combinations of these single specialized models 8-10, and full feature models 12-13, that are trained on RSSI and frequency (except for BLE) for all signal types}
	\begin{scriptsize}
	\begin{tabular}{r|l}
		No. & Model\\
		\hline
		1 & BLE threshold based (GAEN approach)\\
		2 & BLE Decision tree (DT)\\
		3 & BLE Random Forest (RF)\\
		4 & IEEE 802.11 2.4 GHz DT\\
		5 & IEEE 802.11 2.4 GHz RF\\
		6 & IEEE 802.11 5 GHz DT\\
		7 & IEEE 802.11 5 GHz RF\\
		8 & Unweighted combined special DT\\
		9 & Weighted combined special DT\\
		10 & Unweighted combined special RF\\
		11 & Weighted combined special RF\\
		12 & Combined general DT\\
		13 & Combined general RF
	\end{tabular}
\end{scriptsize}
	\label{tbl:models}
\end{table}
In order to make the combined special approach clearer, we describe model 8 as an example: it is a combination of models 2, 4 and 6. The weighted version of this approach uses the weighting described in equation \ref{weighteq}, using the results from the Decision Tree or Random Forest single signal model. These weighting factors are chosen to consider both bands, the two 2.4~GHz signals and one 5~GHz signal, equally.

	\begin{equation}\label{weighteq}
		0.25 \text{ BLE} + 0.25 \text{ WiFi 2.4 GHz} + 0.5 \text{ WiFi 5 GHz}
	\end{equation}

\section{Data set}
\label{section:dataset}
To the best of our knowledge, there is no publicly available data set that contains signal strength values per distance for \ac{BLE} and IEEE 802.11 signals. Therefore, we captured such a data set ourselves in various environments. In this section, we present information on the data set. The data set is publicly available on GitHub\footnote{\url{https://github.com/elanfer/multi-channel-paper-data}}.

\subsection{Data collection setup}
\begin{figure}[tb]
	\subfloat[Ground truth collection\label{gt_setup}]{\includegraphics[height=0.11\textheight]{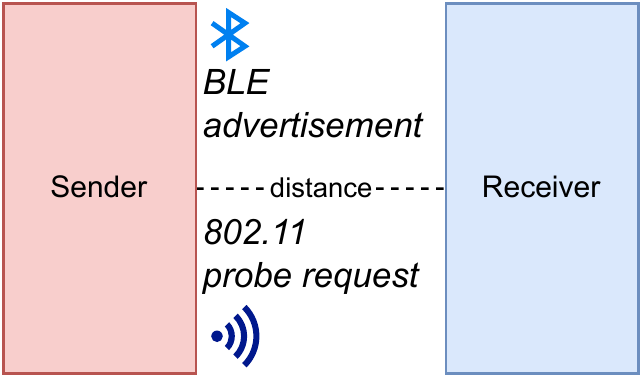}}
	\qquad
	\subfloat[Scenario measurements\label{scene_setup}]{\includegraphics[height=0.11\textheight]{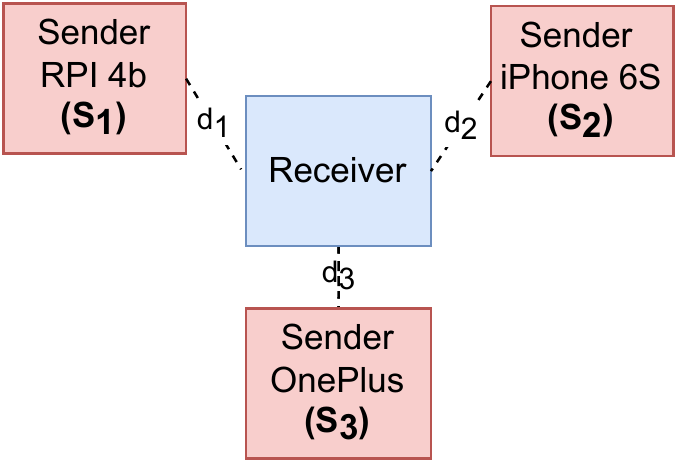}}
	\caption{Setup for measurements}
	\label{fig}
\end{figure}

In order to collect data, we utilized two measurement setups. One for measuring ground truth data, visualized in Figure~\ref{gt_setup}, and a second one to measure real-life scenarios, presented in Figure~\ref{scene_setup}. The main difference between the two setups is that in the ground truth setup only one device is sending out signals, with a constant device orientation over all iterations. In the scenario measurements, these could be multiple devices with varying orientations.

We use three different devices for transmission. These devices are a \emph{Raspberry Pi 4b}, a \emph{OnePlus Nord N10 5G}, and an \emph{Apple iPhone 6S}. With this selection we want to represent the differences between Android and iOS devices. On the two smartphones we use the German Corona-Warn-App to generate \ac{BLE} advertisements. On the Raspberry the advertisements are sent using the \emph{bluez sample advertising script}\footnote{\url{https://github.com/bluez/bluez/blob/master/test/example-advertisement}}. The 802.11 probe requests are sent by triggering network scanning regularly. On the receiver side, which is a Raspberry 4b too, the wireless interface is put into monitor mode using the \emph{nexmon} kernel patch\footnote{\url{https://github.com/seemoo-lab/nexmon}}, to sniff the 802.11 probe requests and a sniffer script for BLE advertisements is developed based on the Rust \emph{bluez} crate\footnote{\url{https://github.com/laptou/bluez-rs}}. 

\subsection{Measurement environments}
\begin{figure}[bt]
	\subfloat[Exterior view\label{1a}]{\includegraphics[height=0.21\textheight]{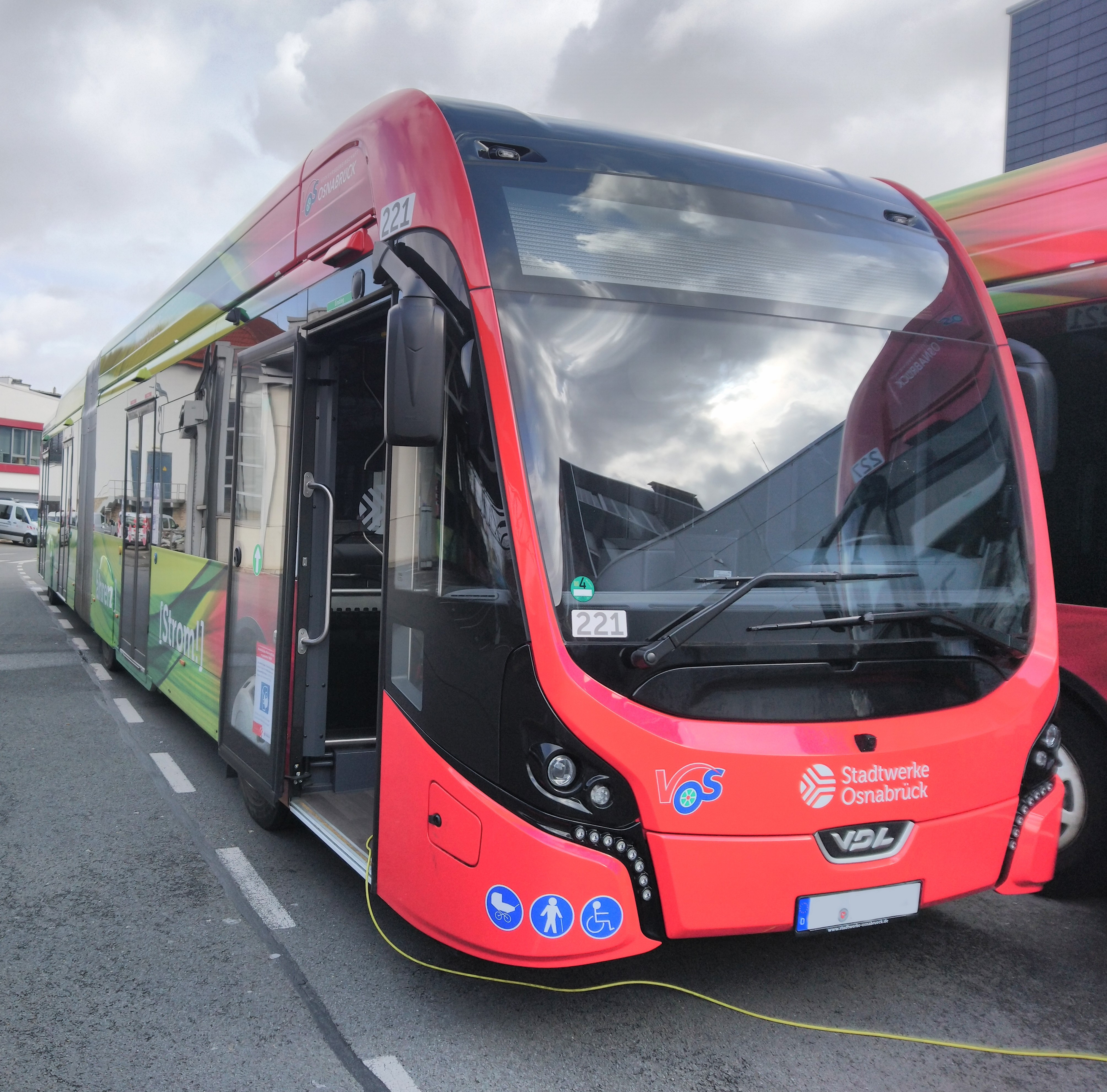}}
	\hfill
	\subfloat[On-board setup\label{1b}]{\includegraphics[height=0.21\textheight]{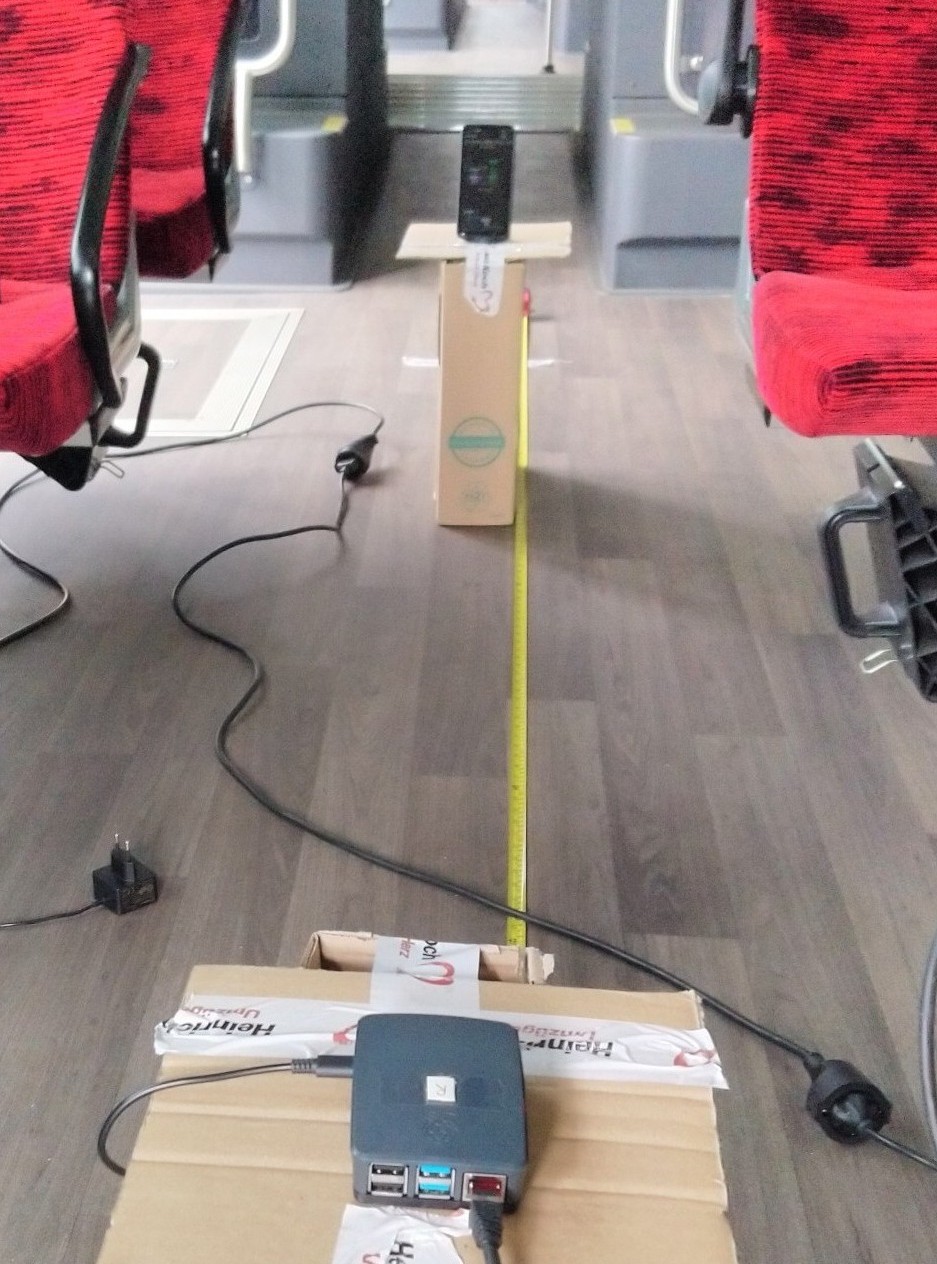}}
	\caption{Bus measurement environment}
	\label{images:bus}
\end{figure}

The measurements were carried out in the following four environments:
\begin{enumerate}
	\item Office room, with a size of $26$~m$^2$
	\item An articulated public transport bus
	\item Outdoors in an empty parking lot, being at least $3$~m away from the next building
	\item On a train, in wagon with open compartments (Deutsche Bahn Intercity Train)
\end{enumerate}
Except for the train environment, we capture data with both the ground truth setup and the scenario setup. For the train environment only scenario measurements is taken. The data is collected during a normal journey. We had no exclusive wagon just for the measurements. However, since we had a very realistic environment, with other devices from passengers that also sent Bluetooth signals, we include these measurements in our data set, too. Moreover, we measured the exact distance of our deployed devices. In the other three environments, we measure at least 15 minutes per distance using the ground truth setup. The distances under measurement are in the range of $50$~cm to $400$~cm and increased by $50$~cm after each iteration. In the bus and outdoor sets, we also measure the distances $500$~cm and $600$~cm to have more than two distances in the \emph{safe} category. The office space is too small to accommodate these distances as well. We show an example of our bus measurement setup in Figure~\ref{images:bus}. While measuring the scenario sets we used different combinations where the devices are placed on seats, to establish realistic scenarios of a bus journey, for example. Additional information on the scenarios is provided in the data set repository.

\subsection{Calibration}
We need to perform a calibration, as the thresholds used by \ac{GAEN} are based on attenuation and not on a device-specific RSSI. In the actual implementation a list of calibration factors by Google is used. This list is based on a calibration procedure designed by Google \cite{Exposure92:online}. In our study, we used a \emph{Raspberry Pi 4b}, a \emph{OnePlus Nord N10 5G}, and an \emph{Apple iPhone 6S}. For these devices no calibration values are available in the list, and the calibration method of Google is not usable, because it is based on a \emph{Pixel 4} device, which is not available for our experiments. To overcome this issue, we took the average RSSI per distance for all devices in all environments we measured, and used these values in a linear program. The program is implemented with the python package \emph{ortools}\footnote{\url{https://developers.google.com/optimization}}. We use constraints, in a way that the \emph{very close} distances have to stay below the threshold of $55$~dB. We relax the variables for the other two distance classes to get as close as possible to the threshold values and to keep the program solvable. This yields a better fit of the calibration to the \emph{very close} distances than to the other two classes. The resulting correction factors are listed in equation \ref{calibration}.
\begin{equation}\label{calibration}
	\begin{aligned}
		\text{iPhone}_\text{corr}:= 16.92,~\text{OnePlus}_\text{corr}:= -1.98,~\text{Pi}_\text{corr}:= 19.66
	\end{aligned}
\end{equation}

\section{Evaluation}
\begin{figure*}[tb]
	\centerline{\includegraphics[width=\linewidth]{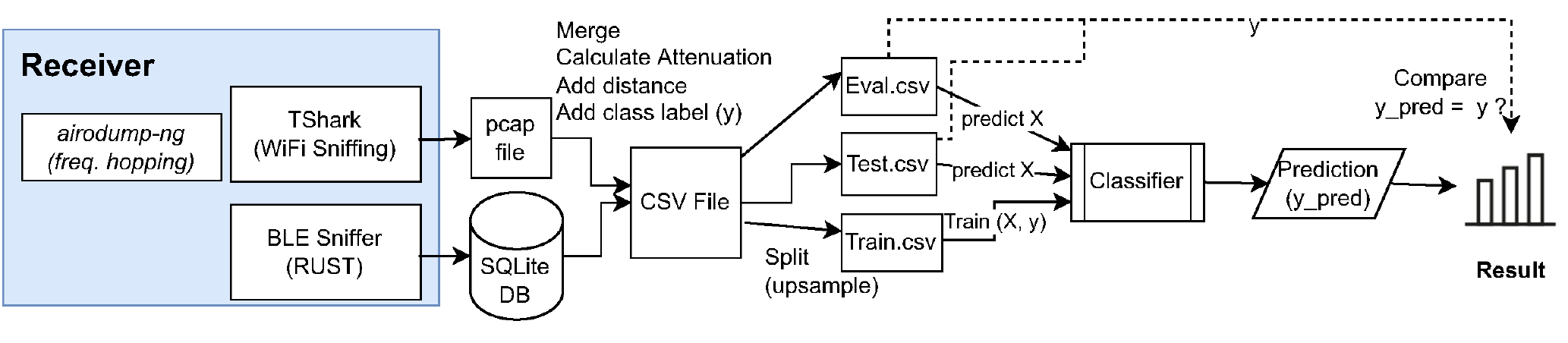}}
	\caption{Data flow overview for the entire setup used in this paper, starting with the recording of the signals and ending with the evaluation of the classifiers}
	\label{img:dataflow}
\end{figure*}

\label{section:evaluation}
In this section, we present the evaluation of our hypothesis. We utilize our novel data set containing \ac{BLE} and IEEE~802.11 data. Before evaluation, we perform a preprocessing step on the data to balance between distances and environments, and to split the data into training, testing, and evaluation data. The preprocessing and the data flow will be briefly described first. Next, we discuss the selection of the metrics used for evaluating the models. Finally, we present the results, with insights into the ground truth data, such as a comparison of all three signal types and the classification performance using the \ac{GAEN} approach with thresholds of the German implementation. 

\subsection{Data processing}
The number of samples per data set is not equal, due to varying distances measured using the scenario setup. Thus, a balancing of the data is needed to avoid overfitting. The overall data flow that is used to capture the samples up to the classifier and the evaluation is illustrated in Figure~\ref{img:dataflow}. In the first step, the recorded \ac{BLE} and IEEE~802.11 data needs to be merged. In both sets, traces of random MAC addresses need to be identified to match the data, the data is available in single measurement runs, that are used for the matching process. This ensures stable distances for matching the three different signals. We use a timing based approach on the \ac{BLE} data to identify the new MAC address after an address and payload roll. All previously seen MAC addresses before the roll and all seen 30 seconds after the roll are filtered, resulting in most cases in only a single address, which is the new MAC address. In collision cases, when two devices roll at the same time, we use the mean RSSI to identify which address belongs to the address under inspection. Using this method we are able to find all packets sent out by a certain device using \ac{GAEN}. For the IEEE~802.11 probe requests, we use a fingerprinting-based approach. We identify the devices based on their features and transmission rates supported, sent in the probe request frames \cite{10.1145/2897845.2897883}. Due to the small number of devices in our recorded data, this approach is very accurate, as comparison to a manual MAC address trace extraction shows. Next, we match \ac{BLE} packets with IEEE~802.11 probe requests, based on the time sent, resulting in a CSV file containing both data sources. Then, the data for each distance in each environment is upsampled, having every class label (\emph{very close}, \emph{close}, and \emph{safe}) and environment equally represented in the set. From this set, a random sample for each class label of $100,000$ is taken, stratified by distances. This overall set is split into a training, test, and evaluation set. The training set consists of $60\%$ of the samples, and the test and evaluation set of $20\%$ each (cf.\ \cite{ng2016nuts}). These sets are then used for training, testing, and evaluating the various machine learning models. From the upsampled data set, except for the train data, we leave data captured in measurements using the \emph{scenario setup} (multiple devices sending in real-life like scenarios) out. These raw samples are later used to also evaluate the classifiers against completely unseen samples.

\subsection{Metrics}
We use common metrics from the field of machine learning, such as $F_1$-score, precision, recall, accuracy, and confusion matrices, to evaluate the performance of our models \cite{alpaydin2020introduction}. For better readability, we do not present a confusion matrix for every classifier. Instead, we sum up the accuracies and the $F_1$-scores for each class in tables.


\subsection{Ground truth data for BLE and IEEE~802.11}
\begin{figure}[tb]
	\centering
	\includegraphics[width=0.9\linewidth]{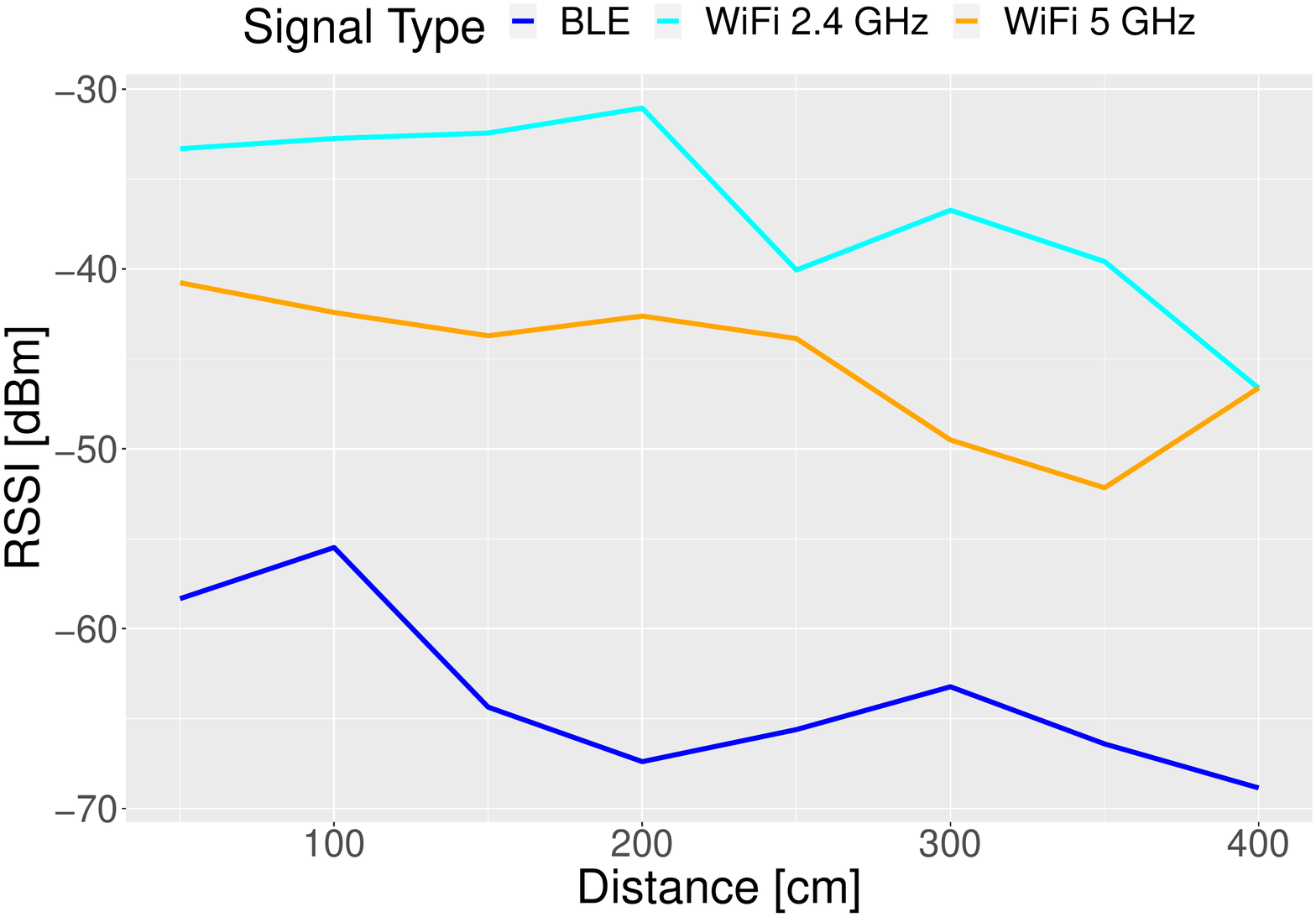}
	\caption{Comparison of BLE and IEEE 802.11 RSSI values per distance, measured in the meeting ground truth setup, using the OnePlus for sending}
	\label{plot:wifiBleComparison}
\end{figure}

As an example, Figure~\ref{plot:wifiBleComparison} shows the ground truth data measured in the meeting room using the OnePlus for sending. For a better overview we connected the average values per $50$~cm in a line plot. As described in the signal propagation background (\ref{section:signalprop}), the BLE signal is affected more strongly by fades, e.g.\ caused by ground reflection.
In comparison, for the 2.4~GHz Wi-Fi signals a correlation of RSSI and distance can be derived, unlike for \ac{BLE}. The 5~GHz signals seem to be less distorted. There is only a single stronger outlier at $400$~cm. Evaluating the \ac{BLE} threshold based approach, which is also used in GAEN, on our data, the distance prediction performance is equally poor as reported in other studies \cite{leith2020coronavirus}.
A confusion matrix for the OnePlus is shown in Table~\ref{table:BLEThreshold:OnePlus}. The accuracy of this method is almost at $0.6$, and worse using other senders (cf. Table~\ref{tab:result-overview}). E.g., for the iPhone, more than every second sample is falsely classified. From the clearer RSSI trends for the IEEE~802.11 signals, on both bands, shown in Figure~\ref{plot:wifiBleComparison}, we expect a better performance and better base for estimating distance. 

\subsection{Classification performances}
The complete result of the model evaluation on our data set is listed in Table~\ref{tab:result-overview}. Our evaluation shows that the Random Forest models, using only a single signal type, performed similar to the Decision Trees. Moreover, the models that combined three single classifiers (models 8-11) did not perform better than a single classifier. The weighted version's performance is comparable to that of the 5~GHz model, but in most cases a few points worse. Our assumption that IEEE~802.11 signals will form a better base for distance classification, appears to hold true. These models performed better than the two \ac{BLE} models in every case. Apart from the clearer trend of RSSI vs.\ distance, the IEEE~802.11 models also utilized the frequency as a feature. Thus, better differentiations are possible. Visualizations of the Decision Trees show that having this information brings good information gain and enables the model to perform better splits. The clear winners are the two models that utilize all available features/signal types in a single classifier, with the Random Forests slightly outperforming the Decision Trees. Overall, regardless of the sender, our multi-channel approach delivers excellent results in the evaluation with our ground-truth data set. Even if the evaluation subset has not been seen by the models before, the data set can be relatively homogeneous due to upsampling. To ensure that the models do not overfit, we performed additional evaluations with our measurement data from the scenario measurements. We sum up the evaluation data for all data sets in Table~\ref{tab:scenario_eval}. Except for the parking lot set using the Raspberry Pi for sending, our multi-channel approach outperforms the threshold-based \ac{BLE}-only approach.  The Raspberry Pi outlier may be explained by the setup, where the Raspberry Pi is the only device that is on the backside of the receiver. Moreover, most of our training data is measured in indoor environments (which is one of the main use cases for contact tracing). This results in a slight specialization of the model.  Generally, our approach improves the classification performance on average by more than $0.3$, regardless of the device that is used for sending.

\begin{table}[t]
	\caption{Confusion matrix only using BLE Threshold on the ground truth data set with the OnePlus used for sending}
	\centering
	\begin{scriptsize}
	\begin{tabular}{|l|l|l|l||l|}
		\hline
		\multicolumn{5}{|c|}{OnePlus Nord N10 5G}\\
		\hline
		true/pred& \textbf{very close} & \textbf{close} & \textbf{safe} & \textbf{$n$-true}\\
		\hline
		\textbf{very close} & \cellcolor{green!5}$12393$ & $6715$ & $891$ & $19999$ \\
		\hline
		\textbf{close} & $5651$ & \cellcolor{green!5}$8433$ & $5916$ & $20000$ \\
		\hline
		\textbf{safe} & $1842$ & $4326$ & \cellcolor{green!5}$13833$ & $20001$ \\
		\hline 
		\hline
		\textbf{$n$-pred} & $19886$ & $19474$ & $20640$ & $60000$ \\
		\hline 		
	\end{tabular}
	\end{scriptsize}
	\label{table:BLEThreshold:OnePlus}
\end{table}

\begin{table*}[]	
	\caption{Classifier overview for the ground truth evaluation set}
	\begin{scriptsize}
	\begin{tabular}{|l|l|l|l|l||l|l|l|l||l|l|l|l|}
		\hline
		\textbf{Model/Metric} &
		\textbf{\begin{tabular}[c]{@{}l@{}}$F_1$-score\\ very close\end{tabular}} &
		\textbf{\begin{tabular}[c]{@{}l@{}}$F_1$-score\\ close\end{tabular}} &
		\textbf{\begin{tabular}[c]{@{}l@{}}$F_1$-score\\ safe\end{tabular}} &
		\textbf{acc.} &
		\textbf{\begin{tabular}[c]{@{}l@{}}$F_1$-score\\ very close\end{tabular}} &
		\textbf{\begin{tabular}[c]{@{}l@{}}$F_1$-score\\ close\end{tabular}} &
		\textbf{\begin{tabular}[c]{@{}l@{}}$F_1$-score\\ safe\end{tabular}} &
		\textbf{acc.}  &
		\textbf{\begin{tabular}[c]{@{}l@{}}$F_1$-score\\ very close\end{tabular}} &
		\textbf{\begin{tabular}[c]{@{}l@{}}$F_1$-score\\ close\end{tabular}} &
		\textbf{\begin{tabular}[c]{@{}l@{}}$F_1$-score\\ safe\end{tabular}} &
		\textbf{acc.} 
		
		           \\ \hline
		\multicolumn{5}{|c||}{OnePlus Nord N10 5G}                                                                                                                                                                                                                                                                                                                                                                                                                                                                                                                                                                                                                                           & \multicolumn{4}{c||}{Apple iPhone 6S}                            & \multicolumn{4}{c|}{Raspberry Pi 4b}                                                                                                                                                                                                                                                                                                                                                                                                                                                                                                                                                                                 \\ \hline
		\multicolumn{1}{|l|}{BLE thresholds}                         & \multicolumn{1}{l|}{0.62}                                                                                                                                                               & \multicolumn{1}{l|}{0.43}                                                                                                                                                          & \multicolumn{1}{l|}{0.68}                                                                                                                                                         & 0.58                                              & 0.68                                                                                                                                                               & 0.43                                                                                                                                                          & 0.31                                                                                                                                                         & 0.49                                              & \multicolumn{1}{l|}{0.55}                                                                                                                                                               & \multicolumn{1}{l|}{0.36}                                                                                                                                                          & \multicolumn{1}{l|}{0.39}                                                                                                                                                         & 0.43                                              \\ \hline
		\multicolumn{1}{|l|}{BLE DT}                                 & \multicolumn{1}{l|}{0.69}                                                                                                                                                               & \multicolumn{1}{l|}{0.55}                                                                                                                                                          & \multicolumn{1}{l|}{0.7}                                                                                                                                                          & 0.65                                              & 0.74                                                                                                                                                               & 0.41                                                                                                                                                          & 0.68                                                                                                                                                         & 0.63                                              & \multicolumn{1}{l|}{0.65}                                                                                                                                                               & \multicolumn{1}{l|}{0.52}                                                                                                                                                          & \multicolumn{1}{l|}{0.61}                                                                                                                                                         & 0.6                                               \\ \hline
		\multicolumn{1}{|l|}{WiFi 2.4 GHz DT}                        & \multicolumn{1}{l|}{0.8}                                                                                                                                                                & \multicolumn{1}{l|}{0.69}                                                                                                                                                          & \multicolumn{1}{l|}{0.8}                                                                                                                                                          & 0.76                                              & 0.81                                                                                                                                                               & 0.73                                                                                                                                                          & 0.75                                                                                                                                                         & 0.76                                              & \multicolumn{1}{l|}{0.78}                                                                                                                                                               & \multicolumn{1}{l|}{0.72}                                                                                                                                                          & \multicolumn{1}{l|}{0.75}                                                                                                                                                         & 0.75                                              \\ \hline
		\multicolumn{1}{|l|}{WiFi 5 GHz DT}                          & \multicolumn{1}{l|}{0.76}                                                                                                                                                               & \multicolumn{1}{l|}{0.64}                                                                                                                                                          & \multicolumn{1}{l|}{0.42}                                                                                                                                                         & 0.67                                              & 0.82                                                                                                                                                               & 0.8                                                                                                                                                           & 0.86                                                                                                                                                         & 0.82                                              & \multicolumn{1}{l|}{0.94}                                                                                                                                                               & \multicolumn{1}{l|}{0.83}                                                                                                                                                          & \multicolumn{1}{l|}{0.85}                                                                                                                                                         & 0.87                                              \\ \hline
		\multicolumn{1}{|l|}{Comb.  FF DT}                           & \multicolumn{1}{l|}{0.98}                                                                                                                                                               & \multicolumn{1}{l|}{0.98}                                                                                                                                                          & \multicolumn{1}{l|}{0.99}                                                                                                                                                         & 0.98                                              & \cellcolor[HTML]{ADFFA6}0.99                                                                                                            & 0.98                                                                                                                                                          & 0.98                                                                                                                                                         & 0.98                                              & \multicolumn{1}{l|}{0.99}                                                                                                                                                               & \multicolumn{1}{l|}{0.98}                                                                                                                                                          & \multicolumn{1}{l|}{0.98}                                                                                                                                                         & 0.98                                              \\ \hline
		\multicolumn{1}{|l|}{Comb. FF RF}                            &
		\cellcolor[HTML]{ADFFA6}0.99 & \cellcolor[HTML]{ADFFA6}0.99 & \cellcolor[HTML]{ADFFA6}1.00 & \cellcolor[HTML]{ADFFA6}0.99 &
		 \cellcolor[HTML]{ADFFA6}0.99 & \cellcolor[HTML]{ADFFA6}0.99 & \cellcolor[HTML]{ADFFA6}0.99 & \cellcolor[HTML]{ADFFA6}0.99 &
		  \cellcolor[HTML]{ADFFA6}1.00 & \cellcolor[HTML]{ADFFA6}0.99 & \cellcolor[HTML]{ADFFA6}0.99 & \cellcolor[HTML]{ADFFA6}0.99 \\ \hline
	
	\end{tabular}
\end{scriptsize}
\label{tab:result-overview}
\end{table*}

\begin{table*}[tb]
	\caption{Model performance overview showing the macro-$F_1$ for all sets per device}	
	\centering
	\begin{scriptsize}
		\begin{tabular}{|l|l|l|l||l|l|l||l|l|l|}
			\hline
			Set/Model                         & \textbf{BLE Thresh.} & \textbf{Comb. FF RF} & \textbf{delta}                    & \textbf{BLE Thresh.} & \textbf{Comb. FF RF} & \textbf{delta}                    & \textbf{BLE Thresh.} & \textbf{Comb. FF RF} & \textbf{delta}                    \\\hline
			\multicolumn{4}{|c||}{One Plus Nord 10 5G}                                                                                                                                      & \multicolumn{3}{c||}{Apple iPhone 6S}                                                                                                      & \multicolumn{3}{c|}{Raspberry Pi 4b}                                                                                                      \\\hline
			\textbf{GT Set}  & 0.58                                        & 0.99                                  & \cellcolor[HTML]{ADFFA6}+0.41 & 0.47                                        & 0.99                                  & \cellcolor[HTML]{ADFFA6}+0.52 & 0.43                                        & 0.99                                  & \cellcolor[HTML]{ADFFA6}+0.56 \\\hline
			\textbf{Parking} & 0.21                                        & 0.60                                  & \cellcolor[HTML]{ADFFA6}+0.39 & 0.56                                        & 0.68                                  & \cellcolor[HTML]{ADFFA6}+0.12 & 0.58                                        & 0.35                                  & \cellcolor[HTML]{F38E8E}-0.23 \\\hline
			\textbf{Meeting} & 0.50                                        & 0.89                                  & \cellcolor[HTML]{ADFFA6}+0.39 & 0.31                                        & 0.85                                  & \cellcolor[HTML]{ADFFA6}+0.54 & 0.19                                        & 0.91                                  & \cellcolor[HTML]{ADFFA6}+0.72 \\\hline
			\textbf{Bus}     & 0.49                                        & 0.61                                  & \cellcolor[HTML]{ADFFA6}+0.12 & 0.60                                        & 0.72                                  & \cellcolor[HTML]{ADFFA6}+0.12 & 0.26                                        & 0.47                                  & \cellcolor[HTML]{ADFFA6}+0.21 \\\hline
			\textbf{Total}   & 0.44                                        & 0.77                                  & \cellcolor[HTML]{ADFFA6}+0.33 & 0.49                                        & 0.81                                  & \cellcolor[HTML]{ADFFA6}+0.32 & 0.36                                        & 0.68                                  & \cellcolor[HTML]{ADFFA6}+0.32\\\hline
		\end{tabular}
	\end{scriptsize}
	\label{tab:scenario_eval}
\end{table*}

\section{Conclusions \& Future work}
\label{section:conclusion}
In this paper, we presented a method to improve the accuracy of the distance classification used in privacy-preserving contact tracing systems. Our own measurements confirm results from other studies that showed using \ac{BLE} signals with thresholds for proximity classification is an error-prone approach. To overcome this issue we proposed an approach of adding signal strength information of IEEE~802.11 signals, for both the 2.4~GHz and 5~GHz bands, to perform a more accurate proximity classification. Our approach outperformed the \ac{GAEN}, \ac{BLE} threshold based, approach significantly. We proved this for three different devices in several environments. This leads to the conclusion that the availability of signal strength information from different channels allows better distance classification.

In our study, we used IEEE~802.11 probe request frames as broadcast messages on IEEE~802.11 in order to combine them with the \ac{BLE} advertisements. We used these packets to achieve an easy deployment. Unfortunately, there are several limitations when using such probe requests. For example, there are many cases where these messages are not sent, which results in a difficult mapping of the signal types. Moreover, when a smartphone is connected to a network, the monitor mode can not be used, otherwise the connection needs to be interrupted. In addition, probe requests contain information that make it easier to identify a device or user. The combination of probe requests and \ac{BLE} advertisements, will result in an increased risk to user privacy. To overcome this privacy risk, we propose the use of a customized data frame. Such a data frame should contain the same payload that is used for the \ac{BLE} advertisements, so essentially the \ac{RPI}. This new custom frame would also solve the issue of matching \ac{BLE} signals with IEEE~802.11 signals.  Such frames need to be sent immediately after or before sending the \ac{BLE} signal, to ensure that both signals hold a stable distance to the receiver. This is especially important in moving scenarios.

Moreover, in future work, a calibration method for IEEE~802.11 signals is needed. This will enable the training of just a single machine learning model instead of device-specific models as we them. Another issue is power consumption. Periodically sending and receiving IEEE~802.11 signals consumes more power than doing this with \ac{BLE} only. One way of solving this could be the utilization of an opportunistic sending approach, such as the Trickle algorithm \cite{levis2004trickle}. This would save battery power, but also enables a better proximity classification when other clients are detected via the \ac{BLE} advertisements. One last point to mention is the issue of device orientation, occlusions, and movements. This can highly influence the signal strength. Therefore, a promising way could be the utilization of \ac{IMU} or gyro sensor data, to detect such situations, to apply correction factors.

\begin{acronym}
	\acro{BLE}{Bluetooth Low Energy}
	\acro{DP-3T}{Decentralized Privacy-Preserving Proximity Tracing}
	\acro{GAEN}{Google/Apple Exposure Notification}
	\acro{IMU}{Inertial measurement unit}
	\acro{LOS}{line of sight}
	\acro{PDU}{Protocol Data Unit}
	\acro{RSSI}{Received Signal Strength Indicator}
	\acro{RPI}{Rolling Proximity Identifier}
	\acro{TEK}{Temporary Exposure Key}
\end{acronym}
\bstctlcite{IEEEexample:BSTcontrol}
\bibliographystyle{IEEEtranS}
\bibliography{mybib.bib}

\end{document}